\documentclass[12pt,preprint]{aastex}

\def\mj{M$_{\rm J}\ $}
\def\rj{R$_{\rm J}\ $}
\def\etal{{et~al.\,}}

\def\ro{R$_\odot$}

\def\mp{M$_{\rm p}$}

\def\teff{T$_{\rm eff}\,$}

\def\sles{\lower2pt\hbox{$\buildrel {\scriptstyle <}
   \over {\scriptstyle\sim}$}}
\def\sgreat{\lower2pt\hbox{$\buildrel {\scriptstyle >}
   \over {\scriptstyle\sim}$}}

\begin{document}


\title{Theoretical Radii of Transiting Giant Planets: The Case of OGLE-TR-56b}

\author{A. Burrows\altaffilmark{1}, I. Hubeny\altaffilmark{1,2}, W.B. Hubbard\altaffilmark{3}, 
D. Sudarsky\altaffilmark{1}, \& J.J. Fortney\altaffilmark{4}}

\altaffiltext{1}{Department of Astronomy and Steward Observatory, 
                 The University of Arizona, Tucson, AZ \ 85721;
                 burrows@zenith.as.arizona.edu, sudarsky@as.arizona.edu}
\altaffiltext{2}{NOAO, Tucson, AZ 85726; hubeny@noao.edu} 
\altaffiltext{3}{Department of Planetary Sciences, Lunar and Planetary Laboratory, 
                 The University of Arizona, Tucson, AZ \ 85721;
                 hubbard@lpl.arizona.edu}
\altaffiltext{4}{NASA Ames Research Center, Mail Stop 245-3, Moffett Field, CA 94035; jfortney@arc.nasa.gov} 

\begin{abstract}

We calculate radius versus age trajectories  
for the photometrically-selected transiting extrasolar giant
planet, OGLE-TR-56b, and find agreement between theory and observation,
without introducing an ad hoc extra source of heat in its core.  The fact 
that the radius of HD209458b seems larger than the radii of the recently
discovered OGLE family of extremely close-in transiting planets 
suggests that HD209458b is anomalous.  Nevertheless, our good fit to OGLE-TR-56b 
bolsters the notion that the generic dependence
of transit radii on stellar irradiation, mass, and age 
is, to within error bars, now quantitatively understood.   

\end{abstract}


\section{Introduction}
\label{intro}

The measurement of the Doppler wobble of more than 120 nearby stars
induced by the presence of a planetary-mass companion has revealed
a population of extrasolar giant planets (EGPs) that is the focus
of an increasing fraction of the world's astronomers
\footnote{see J. Schneider's Extrasolar Planet Encyclopaedia at http://www.obspm.fr/encycl/encycl.html
and the Carnegie/California compilation at http://exoplanets.org 
for current tallies and the associated stellar data.}.  However,
due to the fact that for the vast majority of these EGPs the orbital
inclination ($i$) is not known, only a lower limit, the projected mass (\mp sin(i)),
constrains the actual planetary mass (\mp).  Though the orbital distances ($a$), periods ($P$),
and eccentricities ($e$) are well determined, to study an EGP in physical detail
requires physical information, such as the actual mass, the radius, and the spectrum.

The detection of planetary transits across the face of the 
parent star provides the first two of these desiderata
and with both masses and radii the structural (and to some degree compositional)
character of the EGP can be studied (Guillot et al. 1996).  
HD209458b was the first EGP to be detected to
transit its primary (Henry \etal 2000; Charbonneau 
\etal 2000) and at a distance of only $\sim$47 pc,
it is bright enough to yield (using HST/STIS)
a transit light curve with $\sim$100-{\it micro}magnitude 
precision (Brown \etal 2001).  Proximity also enables precise
radial velocity measurements. As a consequence, 
the data for this radial-velocity-selected
transiting EGP are some of the best we can expect (Brown 
\etal 2001; Mazeh \etal 2000; Cody \& Sasselov 2002). 
Furthermore, the overall transit probability for an EGP in the 
Doppler surveys is very roughly 0.1 (fraction close enough) $\times$
0.1 (fraction near $90^{\circ}$ inclination) = 0.01.  Since of order
100 EGPs have been detected, and the Doppler surveys of nearby stars are approaching
completeness, we can't expect too many more like HD209458b. 

It is in this context that the photometrically-selected transiting EGPs OGLE-TR-56b (Konacki \etal 2003;
Sasselov 2003; Torres \etal 2004), OGLE-TR-113b (Bouchy 
\etal 2004; Konacki \etal 2004), and OGLE-TR-132b (Bouchy \etal 2004)  
should be viewed.  The small subset of the stars in the OGLE galactic survey 
that show periodic photometric dips, but that also survive close scrutiny for
false positives (stellar binarity, confusion, etc.), has the potential to add considerably to 
our knowledge of the Radius-Mass relation for EGPs.  Table 
1 gives relevant stellar and planetary data for
the known transiting systems, along
with associated references.  

However, at distances
of perhaps 1500 pc, even 8-meter class telescopes can't provide the
level of Doppler precision necessary to compete on 
a regular basis with that achievable by the ongoing
radial-velocity surveys in the solar neighborhood.  
Moreover, at a distance of $\sim$1500 pc, an accurate
measurement of the depth of the photometric transit is a major challenge.
Nevertheless, the large volume surveyed by the OGLE team, and the large volumes that
can be surveyed using similar photometric approaches, imply that such programs 
have the potential to yield a rich harvest of transiting EGPs.  
Ground-based photometric transit surveys will pave the way for the more precise space-based
surveys to be conducted by Kepler (Koch et al. 1998) and COROT (Antonello \& Ruiz 2002).
Therefore, we can expect in the years to come a large family of EGPs
for which both radii and masses are known and, hence, for which a robust theory
of EGP radii will be required.  

Theories for the radius of HD209458b in particular (Burrows \etal 2000; Hubbard \etal 2001; 
Fortney \etal 2003; Burrows, Sudarsky, \& Hubbard 2003; Bodenheimer, 
Lin, \& Mardling 2001; Bodenheimer, Laughlin, \& Lin 2003;
Guillot \& Showman 2002; Showman \& Guillot 2002; Allard et al. 2003; Baraffe \etal 2003) and
for irradiated EGPs (``roasters") in general (Guillot \etal 1996; Burrows, Sudarsky, \& Hubbard 2003; 
Baraffe \etal 2003; Chabrier \etal 2004) are appearing that address many of the 
issues that surround theoretical calculations of the radii of irradiated
EGPs and their evolution.  We refer to the discussion in 
Burrows, Sudarsky, \& Hubbard (2003, BSH) for a critique of the literature and a 
summary of the various methods.  

The apparent anomaly
of the OGLE transits is the small inferred transit radii in the optical (Table 1),
given the larger measured radius for HD209458b.  Since the orbital distance
of OGLE-TR-56b in particular is half (0.0225 AU) that of HD209458b (0.045 AU),
it might have been expected that the greater stellar insolation
would have ``expanded" its radius even more than that of HD209458b.  To explain
the large radius of HD209458b, a number of theorists have invoked an additional heat/power
source in the core, due either to the conversion of a fraction of the intercepted 
stellar light into deeply-penetrating mechanical waves (Baraffe \etal 2003; 
Guillot \& Showman 2002; Showman \& Guillot 2002), or to the presence
of an as-yet-unseen companion that induces a slight eccentricity in HD209458b
(Bodenheimer, Laughlin, \& Lin 2003).  Such an eccentricity can result in tidal heating.
Chabrier \etal (2004) have calculated models for OGLE-TR-56b, but have suggested
that the injection of added power might be needed to compensate for what would otherwise be
a $\sim$0.1 \rj\footnote{\rj = $7.149\times 10^4$ km, Jupiter's radius} 
discrepancy between their determination and the 
central value of the measured transit radius.   
However, their numbers do clip the lower end of the error bar range.
In this paper, our goal is to explain the measured radius of OGLE-TR-56b using the
tools and approximations described in BSH, without invoking any additional heat source.  
We find that the radius of OGLE-TR-56b can indeed be fitted comfortably using this
theory.

In \S\ref{review}, we summarize our approximations and approach.
In \S\ref{results}, we present our theoretical results for the 
evolution with age of the radius (R$_{p}$) of the transiting planet 
OGLE-TR-56b and compare theory with observation. In \S\ref{conclusion}, we 
review our conclusions and attempt to put them, as well as  
our physical theory, into the broader context of the family
of irradiated EGPs, both those that are known and those to be discovered.
We end with a synopsis of the improvements in the theory of irradiated
EGPs that are necessary to make further progress.

\section{Techniques and Physically-Motivated Assumptions}
\label{review}

The evolution of an EGP in isolation requires an outer boundary condition
that connects radiative losses, gravity ($g$), and core entropy ($S$).
In this case, the radiative losses are given by $4\pi {\rm R}_p^2 \sigma {\rm T}_{eff}^4$,
where \teff is the effective temperature, R$_{p}$ is the planet's radius,  
and $\sigma$ is the Stefan-Boltzmann constant.  When there is no irradiation, 
the effective temperature determines both the flux from the core and from the entire object.
A grid of \teff, $g$, and $S$, derived from detailed atmosphere calculations,
can then be used to evolve the EGP (Burrows \etal 1997).

Stellar irradiation can drastically alter an EGP's atmosphere and the relationship
between the core entropy, gravity, and core luminosity.  The latter
can be tied to an effective temperature (\teff), but this is now very much 
lower than the equilibrium temperature (Sudarsky, Burrows, and Hubeny 2003)
achieved in the roaster's upper atmosphere.  It is this \teff that determines
the rate with which the irradiated planet's core cools (BSH) and it is the core entropy
that dominates the determination of the radius of a planet of a given mass.  Hence,
when there is stellar irradiation, \teff gives the flux from the core and determines the inner
boundary condition for the atmosphere problem, but does not determine the total emergent
flux.  This is given by the sum of the irradiation flux and core flux.  As a result,
a more careful atmosphere calculation, one that penetrates deeply into the convective
zone to Rosseland optical depths of $\sim 10^6$ and pressures of $\sim 10^{3}$ bars, 
is required to establish the boundary conditions necessary for 
evolutionary calculations of severely irradiated EGPs.  
We use a specific variant of the stellar atmosphere code
TLUSTY (Hubeny 1988; Hubeny and Lanz 1995), called
COOLTLUSTY (briefly described in Sudarsky, Burrows,
Hubeny 2003), to calculate T/P profiles and evolutionary 
boundary conditions for irradiated EGPs such as OGLE-TR-56b,
the evolutionary code of Burrows \etal (1997),
the $H/He$ equation of state of Saumon, Chabrier, \& Van Horn (1995),
the opacity library described in Burrows \etal (2001), 
an updated version of the thermochemical
database of Burrows \& Sharp (1999),
and a stellar spectrum for OGLE-TR-56 (with an assumed 
spectral type of G2 V) from Kurucz (1994). 

As BSH have shown, the transit radius is not the standard ``1-bar" pressure 
radius (Lindal \etal 1981), nor the ``$\tau = 2/3$" photospheric radius.
It is the radius at which the optical depth (at a given frequency) along the 
chord perpendicular to the radius vector is of order unity.
As a result, the ratio of the photospheric pressure to 
the ``transit pressure" is near ($2\pi$ R$_{p}/H)^{1/2}$,
where $H$ is the pressure scale height (Smith \& Hunten 1990).  This adds $\sim$5 pressure scale heights
to the $\sim$10 pressure scale heights between the canonical photosphere
and the radiative/convective boundary.  As found in BSH, the upshot 
for HD209458b is an increase of $\sim$10\% in its transit radius.
For this paper, we have calculated the transit pressure for OGLE-TR-56b
using the methodology of Fortney et al. (2003) and find an average
value in the optical of $\sim$20 to 30 millibars for
$f = 0.5$ and $f = 0.25$, respectively.  This translates 
into an increase of $\sim$3-4.5\% in the transit radius
of the more massive OGLE-TR-56b.

To carry out calculations of the evolution of R$_{p}$ with age for a given \mp\
and irradiation regime, we must assume a helium fraction (Y$_{He}$), address
the issue of the possible presence of a rocky core, account for variations in the angle
of incidence of the stellar radiation across the planet's surface, 
and address the issue of the difference between the day- and night-side cooling.
For these calculations, we take Y$_{He} = 0.30$.  This is larger than the Y$_{He}$
expected, but can account for the effect of a rocky core.  As shown by BSH for
HD209458b, a 10-Earth-mass core shrinks the planet by only $\sim$3-4\%.  This is 
similar to the effect of increasing Y$_{He}$ by 0.02.  For the more massive OGLE-TR-56b (Table 1),
the effect of a rocky core is smaller still.  As described in BSH, we have introduced 
the flux parameter $f$ which accounts in approximate fashion for the
variation in incident flux with latitude when using a planar atmosphere code.
A value of $f = 1/2$ assumes that there is little sharing of heat between the day and night sides.
A value of $f = 1/4$ assumes that in the calculation of the $T/P$ profile 
the heat from irradiation is uniformly distributed over the entire sphere.
In this paper, we show the results for both assumptions, but favor $f = 1/4$
to approximately account for what may be significant redistribution to the night side.  

The issue of the value
of $f$ is tightly coupled to the day-night cooling difference.  The \teff  
for the core in each hemisphere depends upon the strong atmospheric circulation
currents that advect heat from the day to the night sides (BSH; Guillot \& Showman 2002;
Showman \& Guillot 2002; Menou \etal 2002; Cho \etal 2003; Burkert \etal 2004).
A three-dimensional radiation/hydrodynamic study or Global-Climate-Model (GCM) is beyond
the scope of this paper.  In lieu of that, we assume as in BSH and as do Baraffe \etal (2003)
that the flux from the core in both hemispheres is the same.  This does not mean that 
the $T/P$ profiles are the same at altitude, only that the flux at depth at the 
radiative/convective boundary is the same.

\section{Results for OGLE-TR-56b: R$_p$ versus Age}
\label{results}

Figure \ref{fig:1} depicts evolutionary trajectories 
of the transit radius R$_{p}$ in the optical versus age
for the $f=1/2$ and $f=1/4$ models of OGLE-TR-56b (gold).  For both models,
Y$_{He}$ = 0.30 and \mp = 1.45 \mj.  Included is the corresponding trajectory 
for an isolated planet with OGLE-TR-56b's characteristics.  
Irradiation is seen to increase R$_{p}$ by $\sim$0.2-0.3 \rj,
depending upon age and $f$.  We have used the theory of Fortney \etal (2003)
with our derived $T/P$ and optical depth profiles to 
calculate a transit pressure level and, hence, the magnitude 
of the ``impact parameter" that is the transit radius.  Despite the larger insolation
flux, OGLE-TR-56b's larger gravity results in a slightly smaller
``atmospheric thickness" effect (3-4.5\%) than for HD209458b ($\sim$10\%).  Superposed on Fig. \ref{fig:1} 
are the OGLE-TR-56b data from Table 1, where the age of OGLE-TR-56b is ascribed 
to Sasselov (2003).  For comparison, Fig. \ref{fig:1} includes two representative models
(black) from BSH for the evolution of HD209458b's transit radius, with Y$_{He}$ = \{0.25,0.30\} and $f = 1/2$. The  
age and R$_{p}$ estimates for HD208458b listed in Table 1  
are plotted.  The lowest 1-$\sigma$ error bar for HD209458b is from Cody \& Sasselov (2002), under
the assumption that the lower estimate of the corresponding stellar radius ($\sim$1.1 \ro) obtains (BSH).

As Fig. \ref{fig:1} indicates, our theoretical curves are quite consistent with the OGLE-TR-56b data.
The higher $f$ gives larger R$_{p}$, but by only \sles4\% after a Gigayear (Gyr).
At a young age of $10^8$ years, R$_{p}$ can be near 1.5 \rj, but it is $\sim$1.2-1.25 \rj 
after 2 Gyrs.  We have calculated trajectories (not shown) for the OGLE-TR-56b irradiation regime,
but for \mp = 1.68 \mj and 1.22 \mj. After $\sim 3\times 10^8$ years,
they are within less than 1\% of that for \mp = 1.45 \mj.  R$_{p}$ is a very weak function
of planet mass, reflecting the general ``$n=1$" polytropic character of EGPs (Burrows 
\etal 1997; Burrows et al. 2001).  Even if \mp\ for OGLE-TR-56b
were 0.7 \mj, R$_{p}$ would be larger by only $\sim$0.05 \rj or $\sim$0.1 \rj for ages of
3 Gyr and 0.1 Gyr, respectively.  On Fig. \ref{fig:1}, the small black arrow on the right
indicates the effect of a 10 Earth-mass rocky core on models for HD209458b.  For OGLE-TR-56b,
at twice the mass and $\sim$2.5 times the gravity, the arrow would be less than half as long.
The weak dependence on age, \mp, Y$_{He}$, and the possible presence of a rocky core
implies that we have in our calculations incorporated the essential physics,
chemistry, and radiative effects necessary to explain the transit radius of OGLE-TR-56b,
without invoking an added power source in the core.  The major effects are the stanching
of core cooling (and the decrease in \teff) by irradiation's effect on the atmospheric $T/P$ 
profile (Burrows \etal 2000; BSH) and the 0.04-0.05 \rj difference due to the 
proper definition of the transit radius (BSH; Baraffe \etal 2003).  For OGLE-TR-56b, the pressure at the 
radiative/convective boundary is between 600 and 900 bars, depending upon true age and $f$.
This is slightly lower than that for HD209458b, reflecting the 
larger mass.  The rate with which the radiative front is now penetrating
OGLE-TR-56b is $\sim$200 bars per Gyr, equivalent to a scant $\sim 3\times 10^{-5}$ \mj per 
Gyr.

\section{Conclusions}
\label{conclusion}

We have shown that our theory, which couples spectral, atmospheric, and evolutionary calculations
in a straightforward manner, can explain the measured transit radius of OGLE-TR-56b.
Our calculations yield values for R$_{p}$ that are $\sim$0.1 \rj higher than those
of Chabrier \etal (2004), which without an extra heat source undershoot by $\sim$10\% the
central value of the OGLE-TR-56b R$_{p}$ measurement.
Since they are using a similarly sophisticated approach,
the source of this difference is unknown.

The larger radius of HD209458b is still problematic, but even it can be fitted without
an ad hoc extra power source, if its true transit radius is at the lower end of the measured range (BSH).
It is important to remember that systematic errors still dominate estimates of R$_{p}$.
Furthermore, not only is OGLE-TR-56b smaller than HD209458b, but so too seem
OGLE-113b and OGLE-132b (however, note the large error 
bars in Table 1).  Curiously, all the OGLE roasters have smaller orbital distances.
This implies that HD209458b is the anomaly,
perhaps due to tidal heating caused by an as-yet-unseen second companion 
(Bodenheimer, Laughlin, \& Lin 2003) or to residual systematic errors. 
Hence, a compelling argument can be made that the transit radii of all the OGLE EGPs
are consistent with a model that does not require any extra heating term
beyond that supplied quite naturally through the standard effects
of irradiation and radiative transfer into the convective core.

\begin{figure}
\epsscale{1.00}
\vspace*{-0.7in}
\plotone{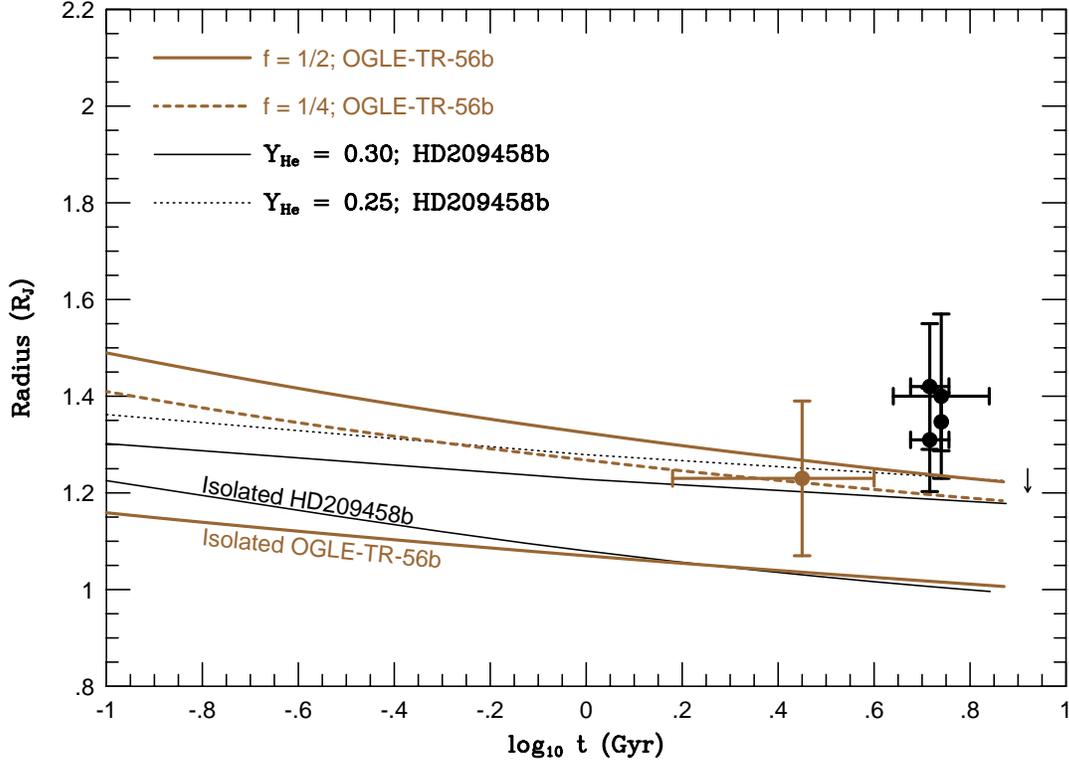}
\vspace*{-0.2in}
\caption{
Theoretical evolutionary trajectories (in gold) of the optical transit radius of
OGLE-TR-56b (in units of \rj) with age (in Gyrs).  The effects of irradiation are included.
A mass of 1.45 \mj, a helium fraction of 0.30, and values of the insolation
parameter $f$ of $1/4$ and $1/2$ are assumed (see BSH and text for discussion).
Model ``Isolated OGLE-TR-56b" is for a 1.45 \mj EGP in isolation.  
The measured optical transit radius and estimated age, accompanied 
by $\pm 1-\sigma$ error bars and taken from Torres \etal (2003) 
and Sasselov (2003), are rendered with the gold cross.  For comparison, evolutionary tracks 
for HD209458b from BSH, assuming helium fractions of 0.25 and 0.30,
along with the corresponding age and radius estimates  
from Mazeh \etal (2000), Brown \etal (2001), and Cody 
and Sasselov (2002), are plotted (all in black).
A model of HD209458b in isolation (``Isolated HD209458b") is also shown.
The short arrow to the right of the HD209458b error boxes depicts the magnitude
of the radius decrease for each 10 Earth-mass increase in the mass of
a possible rocky core in HD209458b.
See text for explanations.
\label{fig:1}}
\end{figure}

The remaining theoretical uncertainties are the actual day-night cooling differences,
the 3D effects of atmospheric circulation and zonal heat transport, and the early
history of the planet.  As shown by Burrows \etal (2000), and as is implied 
on Fig. \ref{fig:1}, if the EGP were born at large orbital distances, but took 
more than $\sim 3\times 10^7$ years to migrate in to its present distance,
then its radius would have shrunk below a value consistent with the measured
R$_{p}$ (for any of the objects listed in Table 1). One could then accommodate
an extra heat source, since it would be needed to compensate for the early
loss of core entropy.  However, such a migration time is deemed rather long,
and we prefer to shave with Occam's Razor.

\acknowledgments

We thank Christopher Sharp, 
Dimitar Sasselov, Adam Showman, Jonathan 
Lunine, Dave Charbonneau, and Drew Milsom for useful discussions
during the course of this investigation. This study 
was supported in part by NASA grants NAG5-10760 and NAG5-13775.
This material is based upon work supported by the National Aeronautics and
Space Administration through the NASA Astrobiology Institute under
Cooperative Agreement No. CAN-02-OSS-02  issued through the Office of Space
Science.

{}

\clearpage

\begin{deluxetable}{llllllll}
\tablewidth{18.2cm}
\tablenum{1}
\tablecaption{Data for Current List of Transiting EGPs\label{EGP_tab}}
\tablehead{
\colhead{EGP} & \colhead{M$_{\ast}$ (M$_{\odot}$)} & \colhead{R$_{\ast}$ (R$_{\odot}$)}
& \colhead{a (AU)}
& \colhead{P (days)} & \colhead{M$_{p}$ (\mj)} & \colhead{R$_p$ (R$_{\rm J}$)} & \colhead{Age (Gyr)}}
\startdata

HD209458b$^1$ & $1.1\pm 0.1$   & $1.2\pm 0.1$  & 0.045 & 3.525 &  $0.69\pm 0.05$ & $1.4\pm 0.17$ & $5.5\pm 1.5$ \\  
{\bf HD209458b$^2$} & $1.06\pm 0.1$   & $1.18\pm 0.1$  & 0.045 & 3.525 &  $0.69\pm 0.02$ & 1.42$^{+0.12}_{-0.13}$ & $5.2\pm 0.5$ \\  
HD209458b$^3$ & $1.1\pm 0.1$   & $1.146\pm 0.05$  & 0.045 & 3.525 &  $\sim 0.69$ & $1.347\pm 0.06$ & - \\  
{\bf OGLE-TR-56b$^4$} & $1.04\pm 0.05$   & $1.1\pm 0.1$  & 0.0225 & 1.212 &  $1.45\pm 0.23$ & $1.23\pm 0.16$ & 2.5$^{+1.5}_{-1.0}$ (S03$^5$) \\  
OGLE-TR-113b$^6$ & $0.77\pm 0.06$   & $0.765\pm 0.025$  & 0.0228 & 1.433 &  $1.35\pm 0.22$ & 1.08$^{+0.07}_{-0.05}$ & - \\  
OGLE-TR-113b$^7$ & $0.79\pm 0.06$   & $0.78\pm 0.06$  & 0.023 & 1.432 &  $1.08\pm 0.28$ & 1.09$\pm 0.10$ & - \\  
OGLE-TR-132b$^6$ & $1.34\pm 0.1$   & 1.41$^{+0.49}_{-0.10}$  & 0.0306 & 1.689 &  $1.01\pm 0.31$ & 1.15$^{+0.80}_{-0.13}$ & - \\  

\tablerefs{ $^1$Mazeh et al. 2000; $^2$Cody \& Sasselov (2002); $^3$Brown et al. (2001);
$^4$Torres et al. (2004); $^5$Sasselov (2003); $^6$Bouchy et al. (2004); $^7$Konacki et al. (2004)}

\enddata
\end{deluxetable}

\clearpage

\end{document}